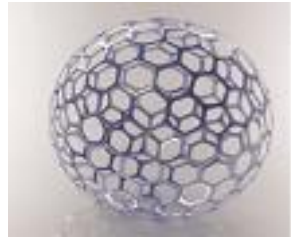

# Title: 'A study of multistage interconnection networks operating with wormhole routing and equipped with multi-lane storage'

## By Eleftherios Stergiou

Associate Professor
Informatics and Telecommunications
University of Ioannina

## Graphical Abstract

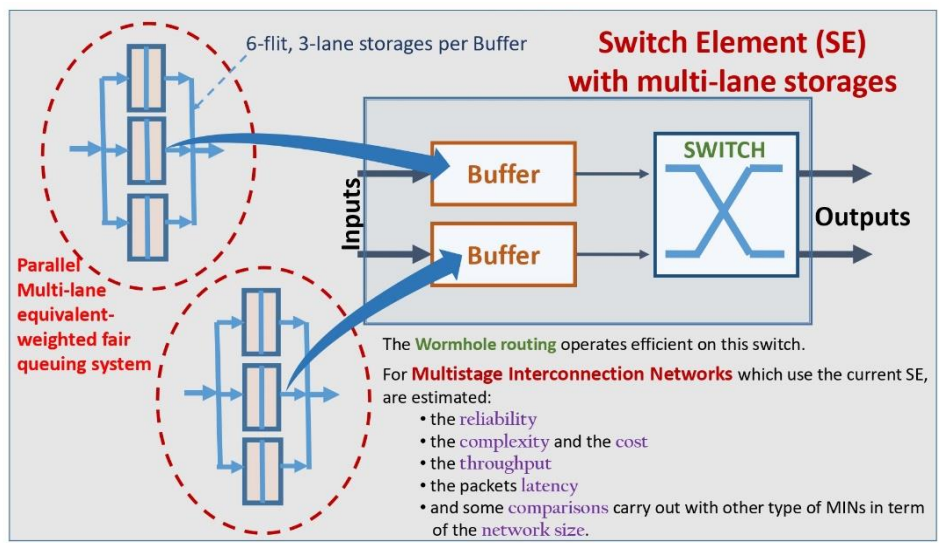

*Switch Elements with multi-lane storages*

# A study of multistage interconnection networks operating with wormhole routing and equipped with multi-lane storage

## Author: Stergiou Eleftherios

**Abstract**

Multistage interconnection networks (MINs) provide critical communication resources between network components with an attractive cost/performance relation. In this paper, a novel architecture for a MIN is proposed. In contrast to other existing networks, this new architecture operates via wormhole routing using a multi-lane equivalent-weighted fair queuing system. Associating the lane storage with each physical channel for movement of the flits allows for high levels of packet flow and makes the system more robust. The resulting flit scheduling schema has been studied thoroughly, and the results are presented in this manuscript. In addition to performance metrics, additional factors such as complexity, cost and reliability are investigated. Since a basic aim of the design of MINs is to achieve a good data flow control mechanism, this proposal is an extremely effective and robust solution. The rationale behind this innovative scheme is to introduce a more efficient network technology, thus providing a better quality of service (i.e. streaming media vs. file transfer).

1. **Introduction**

A multistage interconnection network (MIN) is an apparatus for interconnecting processors in parallel systems, supercomputers and datacentres, and to ensure efficient internetworking [1-3]. The benefits of MINs include their capability to route multiple communication tasks concurrently, as well as their low cost/performance ratio.

*MINs and wormhole routing*

The Banyan type of MIN is a device characterised by the existence of only a single path between each source and destination. Non-banyan interconnection networks are more expensive and more complex to manage. The network-on-chip (NoC) architecture uses interconnection systems to solve the scalability and flexibility communication problems in existing bus technology [4-5]. The problems that arise in such cases are related to the structure of the architecture and the routing choice that is used to ensure efficient data transmission.

The wormhole routing technique [6] seems to be a convenient method that is both appropriate and fully compatible with the operation of MINs, and can improve their speed of data forwarding. A comparison of different types of wormhole routing algorithms is performed in [7]. In addition to MINs and NoCs, wormhole routing algorithms are also implemented in other networks such as mesh networks [8], providing the same advantages. The cost of applying wormhole routing methods to interconnection networks has recently been estimated and reported as not being high [9-10].

There is a continuing interest in developing more efficient MINs, as they are a key aspect of various switching sectors and the requirements for higher data transfer rates are increasing daily.

*This work*

In order to provide high quality of service (QoS) for today's high-speed networks, a new MIN is proposed that uses the wormhole routing method. Multi-lane storage is employed, and these lanes are assigned to packets entering the network (see Figs. 1 and 2). The multi-lane storage arrangement of the buffer was invented to overcome the intense blocking phenomenon that arises when the packet back-pressure technique is used. The flit scheduling scheme that eventually emerges from this proposal appears to be a very promising solution, as it is both an efficient and extremely robust interconnection apparatus.

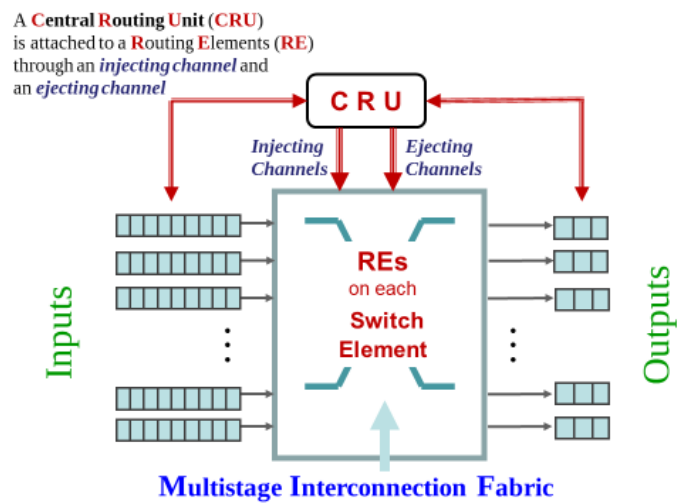

*Figure 1 A general schema for the MIN under study. In addition to the main fabric, the MIN includes a central routing unit (CRU) and routing elements (REs). The CRU is attached to the switching elements (SEs) via REs and injecting/ejecting channels*

In particular, this paper expands upon previous research with regard to the following points:

- A novel MIN architecture is introduced that operates via wormhole routing and uses multi-lane storage buffers in each switch element.
- A ranking of the *complexity* and *cost* factors of the MIN under study compared to other equivalent interconnection apparatuses is presented, in terms of network size and performance.
- The overall point-to-point *reliability* of the MIN is quantitatively assessed. This factor means that the proposed MIN is superior to the other intermediate apparatus.
- The *throughput* metric of the proposed MIN is calculated under various network size configurations; the metric is quantified and useful conclusions extracted that make the behaviour of the MIN.
- The packet *latency* metric is estimated for the current fabric.

The comprehensive approach used here can form a useful tool for evaluating the performance of any finite lane storage buffered MIN, as well as the performance of other similar alternative network architectures.

**Organisation**: The remainder of this paper is organised as follows: Section 2 presents a brief analysis of a multistage network operating with wormhole routing and using divided buffer storage in each node. Section 3 gives basic definitions and an analysis. In Section 4, the reliability of the proposed MIN is analysed and some corresponding numerical results are given for these MINs in terms of network size and the number of lanes used. Section 5 presents a complexity and cost analysis compared to other similar multistage networks. In Section 6, the elements of the simulator construction are described and the method of gathering performance data is discussed. Section 7 presents the results of our simulations, and examines the effect that the buffer division has on the overall network performance. Finally, Section 8 presents the conclusions and anticipated future work.

## 2. Analysis of the proposed MIN

*General description of a MIN*

In general, a typical $N \times N$ MIN with Banyan properties is constructed using $L = Log_k N$ stages of $k \times k$ switching elements (SEs), where $k$ is the degree of the SEs (see Fig. 1, which shows a Delta-type network). Let ($i$) depict an arbitrary number of stages, where ($i$) can take values from 1 to $L$. Generally, each SE consists of $k$ input and $k$ output ports. In this type of fabric, there are exactly ($N/k$) SEs at each stage, giving a total number of SEs for the MIN of $(N/k) \cdot \log_k N$. There are also $(N \cdot \log_k N)$ interconnections between all stages, unlike in the crossbar network, which requires $O(N^2)$ SEs and links (Fig. 2).

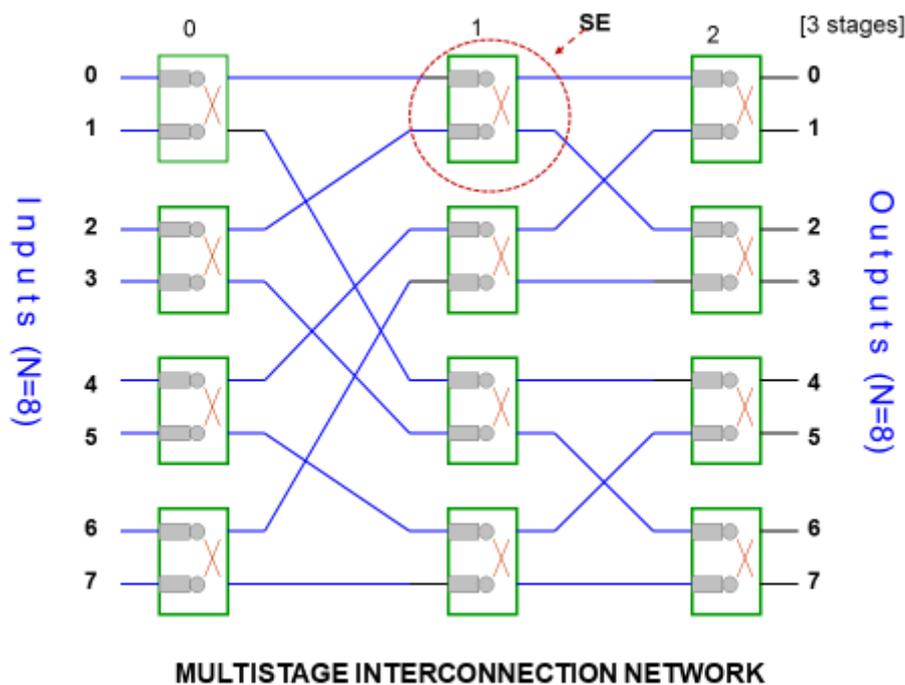

*Figure 2 Typical internal structure of a MIN consisting of 2x2 SEs (N=8 - Delta type)*

Figure 1 shows the general schema of the proposed MIN. The routing schema consists of a *central routing unit* (*CRU*) and individual *routing elements* (*REs*) which are connected to each SE. The CRU is attached to each RE via an *injecting channel* and an *ejecting channel*. Each RE governs its corresponding SE. The injecting and ejecting channels transfer routing information from the REs to the CRU, which has full overall control over routing and implements the whole routing operation. Each SE in the multistage system has a pair of REs, which are the parts of the CRU that send/receive information and control signals. Normally, each SE has one transmission queue (FIFO type) per link, accommodated in one (logical) buffer.

The SEs of the proposed MIN operate under the following assumptions:

(i) A unit of network clock period (time cycle) is divided into two-time sub-periods. In the first sub-period, flow control information is transferred across the network from the last stage to the first. In the second sub-period, packets proceed ahead based on the flow control information.
(ii) The arrival process at each input of the network is calculated based on the assumptions that a packet arrives within a clock cycle with constant probability, and that arrivals are independent of each other (See [21], Definition 7). We denote this probability of arrival by ($\lambda$).
(iii) The proposed MIN operates via wormhole routing. This means that as a packet is created at the source node, it is automatically split into a number of flits. In the next time cycle, the flit scheduling schema starts to operate.
(iv) The creation of packets follows a *binomial distribution*, $bin(k, \lambda/k)$ [21], while the flits of packets generated at each source node are Poissonian, and their destinations are uniformly random.

In the following subsections, the multi-lane buffer organisation used in the proposed MIN and the operation of the wormhole routing are explained.

*Multi-lane organisation of buffers in switch elements*

The buffer of each SE is organised (partitioned) into independent lanes (memory segmentation) that are associated with each input channel, as can be seen in Fig. 3(a).

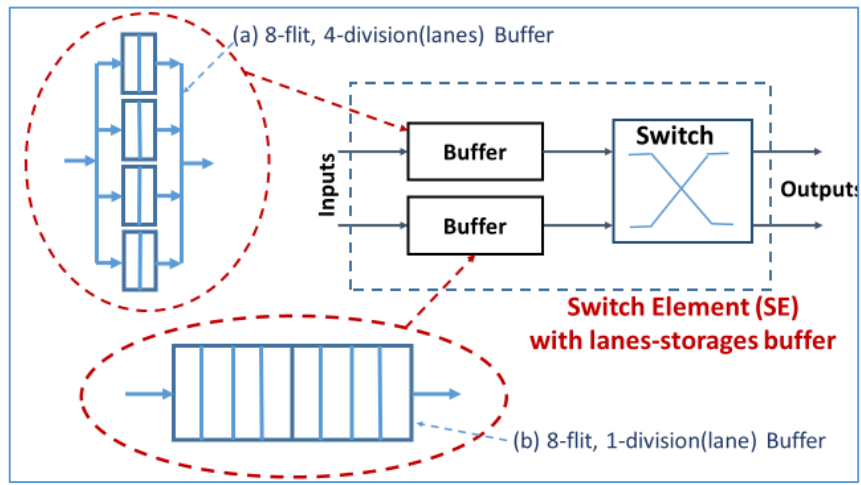

*Figure 3 A switch element organised into independent storage lanes (memory segmentation)*

A multi-lane (or multi-flit) storage scheme is a fair queuing scheduling system that is associated with a corresponding number of virtual channels. Fig. 3 depicts (a) a four-lane storage buffer that is able to store two flits; and (b) a monolithic buffer (single-lane storage) that is able to store eight flits.

As a packet may originate from any input source and is automatically converted into a number of flits, the forwarding process starts in the next time cycle. The header flit is forwarded via lane storage in a snaked manner, and the rest of the flits follow it. The method used to route flits is described in detail in the next paragraph. The parallel multi-lane storage scheme allows parallel data worms to be created, which can better exploit the system's time cycles. This fair queuing scheduling system provides an advantage in terms of the routing process. Although the multi-flit buffer cannot effectively reduce the latency of blocked packets, it can make more channels available when blocking occurs. On the other hand, the main drawback of this architecture is that in order to implement the aforementioned mechanism, more hardware is required for both the CRU and the necessary individual storage lanes, which increases the complexity of the system (Fig. 1).

*Wormhole routing operation*

In a store-and-forward routing mechanism, the buffers of intermediate nodes must be large enough to store packets before they are retransmitted. In all cases in which packets encounter blocking or errors, they must be automatically saved [11-12]. Under this circumstance, the minimum space required is equivalent to the space occupied by at least a packet. In contrast, the proposed MIN operates with a wormhole routing mechanism. In a wormhole switching forwarding method, a packet can be considered as a number of flits [13-14].

Hence, a general overview is given here of the wormhole routing method used in the proposed MIN. The general control of this routing is carried out by the CRU module. The following assumptions are made in the proposed system:

- As the packets originate in the inputs, they are split into flits with a fixed length, and these are snaked along the appropriate route in a pipelined way. As a consequence, the transmission of different packets cannot be interleaved or multiplexed freely over a single physical channel.
- A basic characteristic of multistage fabrics using the wormhole routing technique is that they do not need buffers for whole packets. Unlike the store-and-forward mechanism, wormhole-equipped multistage fabrics use smaller buffers to store one or more flits.
- The header flit, with the relevant routing tag, follows a suitable path in the network. Subsequently, the other flits follow the same path in a pipelined way. The successive buffers and links occupied by the flits of an arbitrary given packet form the wormhole. The length of the worm created here depends on the number of flits into which a packet is divided and the network size of the MIN. Normally, a data worm spans the whole path between the input and output points.
- The routing path is specified by a specific set of internal links over which the packet can be routed.
- When the header flit cannot proceed, due to busy output channels, then the whole flit chain remains in a stagnant state, occupying flit buffers in intermediate SEs. When a worm is not in forward action it may cause blocking to other packets' movement (or better flit blocking) within the fabric.
- All packets arriving at destinations are immediately consumed at the rate of one flit per time cycle; in other words, no blocking is encountered at destination points.
- Wormhole operating fabrics are often equipped with large-scale input buffers.

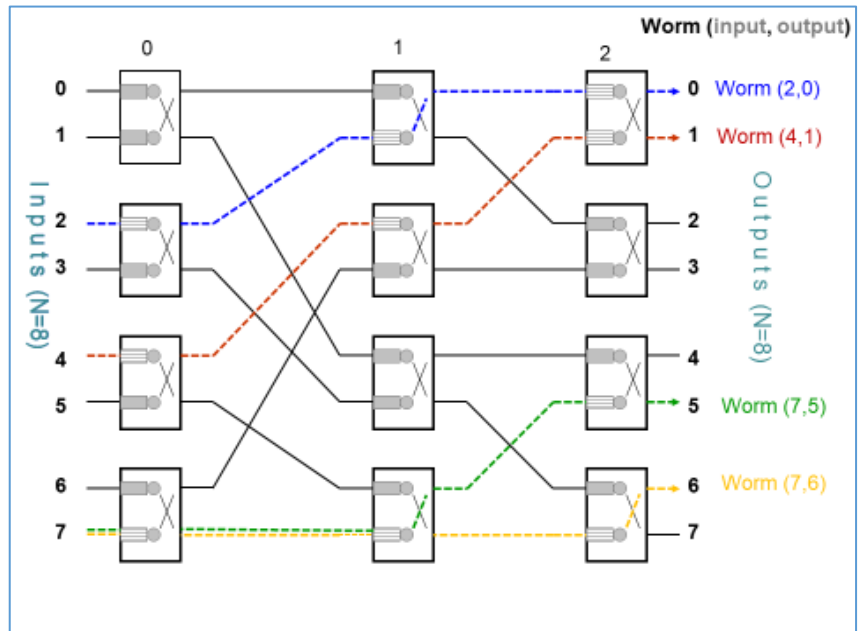

*Figure 4 Example of four worms within a MIN (N=8) with multilane storage buffers*

- The main disadvantage of wormhole routing is the blocking of resources, which arises when pipelines are stalled.
- Since the flit chains of the buffers can easily experience blocking, the system may over time become prone to a snowball phenomenon, meaning that wormhole routing needs to be used within suitable margins and within appropriate limitations.
- The proposed MIN implements storage lanes in each buffer of the SEs. However, flits originating from different packets are not allowed to be mixed within one buffer, since none of the flits (except the header) have identity routing tags. However, by expanding the control logic, one physical channel can be split into several virtual memory lanes. Thus, flits from different packets can be given their own part of the buffer, exploiting a single physical channel medium.
- By selecting the route that a packet takes, the routing algorithm tries to guarantee packet flow and to prevent deadlocks as far as possible.

Finally, the wormhole routing technique provides convenient, cheap, and high-speed routing. Fig. 4 displays a snapshot of a MIN ($N=8$) with multi-lane storage and four worms en route. Wormhole routing is the most common routing mechanism currently used in commercial devices. It is especially susceptible to deadlock conditions, in which two or more packets can block each other indefinitely.

### 3. Basic definitions

This section gives a list of major parameters that are used to describe the proposed MIN and which affect its behaviour.

*Network size/radix x MIN*

A multistage network with $radix\ x$ (where $x$ is equal to $3,4,\ldots,10$) is a MIN with a number of stages $x = L = Log_k N$. The aforementioned radix $x$ depicts MINs with $N = k^x$ inputs/outputs dimension corresponding fabrics.

*The arrival process of packets*

The arrival process of packets at the output queues of the first stage of the network is described by a binomial distribution $bin(k, \lambda/k)$, where $\lambda$ is the fixed probability of a packet being generated by a processor in each cycle. Therefore:

$$x_{k,n}^{(1)} = \begin{cases} \binom{k}{n} \cdot \left(\frac{\lambda}{k}\right)^n \cdot \left(1 - \frac{\lambda}{k}\right)^{k-n}, & \text{for } 0 \leq n \leq k \\ 0, & \text{otherwise} \end{cases} \quad (1)$$

In this case, $k$ is equal to 2, as we select $2 \times 2$ SEs due to the wide use of this configuration in commercial industries. As the packets originate in the inputs of the MIN, the division process of the flits is automatically carried out according to the flit pattern used in the fabric.

*Number of flits ($n_f$) in a packet*

The number of flits ($n_f$) is the number of flits in a data packet, which have a fixed length. The length of each flit must be equal to or less than the size of a storage lane, so that they can be stored in individual storage lanes. In other words, the size of a storage lane must be a multiple of the length of a flit.

*Number of lanes ($n_{lanes}$)*

This indicator shows the number of storage lanes that are applied in each buffer of an SE in the MIN.

*Definition of reliability*

The *reliability* ($r$) of a switch component is defined as the probability of a switch component being operational within a given time period. Each SE component has a different value of reliability. The reliability is denoted by $r_{switch\_n}$, where *switch* signifies the switch type and $n$ depicts the number of gates of the SE. Thus, for example, $r_{SE2x2}$ and $r_{SE2x4}$ represent the reliabilities of $SE_{2x2}$ and $SE_{2x4}$, respectively.

*General reliability analysis of a multi-switch system*

The reliability of a fabric is its ability to overcome unexpected circumstances. In MIN technology, the *point-to-point reliability* (*p-t-p reliability*), also known as the *terminal reliability*, is used. In p-t-p reliability, the probability of at least one fault-free path existing between an input-output pair is considered. The reliability in an arbitrary pair (input-output) is dependent on all the SEs involved, since the failure of each element implies the failure of the current routing action [15-17].

Let us assume a number of n SEs in series, each SE with a reliability of $r_i$; then, the overall reliability of this multi-switch system can be calculated as:

$$R_{nSEs\ in\ Series} = \prod_{i=1}^{n} r_i \quad (2)$$

Assuming a number $n$ of parallel switch components, at least one of these must be active in order for this subsystem to operate successfully. In this case, the overall reliability of the multi-switch system is calculated as:

$$R_{nParallel\ SEs} = 1 - \prod_{i=1}^{n}(1 - r_i) \quad (3)$$

*Buffer utilisation*

In a multi-flit buffer environment, *buffer utilisation* may be a more important parameter than *channel utilisation*. The *normalised buffer utilisation* $u_{N\_buffer}$ is used as an alternative to the traffic load indicator to represent the current amount of internal traffic load in the MIN, and also indirectly shows the traffic margins that a MIN has to exploit (buffers to be full). We define the normalised buffer utilisation as follows:

$$u_{N\_buffer} = \frac{number\ of\ lane-storeges\ in\ busy}{total\ number\ of\ lane-storages\ in\ buffers} \quad (4)$$

In multi-flit buffer environments, buffer utilisation may be more fundamental than channel utilisation.

*Throughput*

The actual value of the *throughput* (number of packets/time cycle) is fundamental metric in these applications. To express the throughput of a MIN using wormhole routing, the *number of flits/time cycle* is usually preferred. These two expressions are equivalent. Different numbers of packets are observed to be circulating within a fabric at different time cycles.

*Maximum throughput* ($Th_{\max}$): The *maximum* (ideal) *throughput* depicts the maximum number of packets ($N_{\max}$) circulating in the network without contention divided by the *ideal average delay* of packets ($\overline{d_{ideal}}$) (without contention). The maximum number of packets in a fabric is in direct correlation with the maximum ideal throughput ($Th_{\max}$). The maximum ideal throughput is defined as:

$$Th = \frac{N_{max}}{\overline{d_{ideal}}} \quad (5)$$

The *ideal average delay* of packets ($\overline{d_{ideal}}$) can be calculated as: $\overline{d_{ideal}} = L + P_{length} - 1$, where $L$ depicts the network *radix* (the absolute distance between source and destination) and $P_{length}$ depicts the length of the packet. The last stage of the MIN does not give rise to packet delay as it is assumed that all packets are absorbed automatically by the next device.

*Normalised throughput* ($Th_N$): Instead of the actual value of throughput, the *normalised throughput* is a preferred metric for comparison purposes. The normalised throughput is defined as:

$$Th_N = \frac{Th_{actual-value}}{Th_{max}} \quad (6)$$

where $Th_{dactual-value}$ is the actual value of throughput from experiment using various values of the network radix and numbers of storage lanes.

*Analysis of packet delay*

The overall packet delay or latency is defined as the time that elapses between a packet being prepared for injection into the network at the source node and the same packet being received at the destination node. When a MIN is used to support packet-passing, the time required to move data between SEs is critical to system performance, as it determines the minimum scaling of data in the data transfer process. Between a packet originating at an arbitrary node ($i$) and reaching its destination, it can be affected by the following types of delay: *waiting delay*, *service delay* and *traverse delay.* The sum of these gives the overall communication packet delay of the fabric.

1) *Waiting delay* ($d_{wait(i,j)}$) at the arbitrary injecting channel ($j$). This is a start-up delay, meaning that the time is incurred at the input node ($i$). This delay is independent of the routing scheme (e.g. *store-and forward* or *wormhole*). This is the time required by the system to handle the packet at the source node, and can be determined based on the time of arrival of the packet and the time at which the packet (the header flit) is initially injected into the channel ($j=1$), i.e. when the service time starts. This value mainly depends on the system design. Since the incoming flits follow a Poisson distribution, the *waiting delay* can be determined using the M/G/1 model.

2) *Service delay* ($d_{serv(i,j)}$) at the arbitrary injecting channel $(j)$. This is the time that elapses between the time slot in which the header flit of the packet is accepted by the injecting channel ($j=1$) to the time slot in which the tail flit of the packet emerges from the injecting channel. In *wormhole* routing, the flits of an arbitrary packet are spread over many links along the packet's path, meaning that the *service delay* ($d_{serv(i,j)}$) at the injecting channel depends on the service time of the subsequent channels. When the header flit is blocked, the following flits of the worm are usually also blocked. In long-worm routing, the service time at the injecting channel includes the waiting times due to blocking at all subsequent channels. Thus, the service delay includes all the individual delays that are encountered during the lifetime of a packet inside in the fabric. These individual delays arise from conflicts over the use of shared resources, for example delays from channel contention in which two packets simultaneously lay claim to the same channel. The service delay is related to the dynamic behaviour of the network resulting from the passing of multiple packets; it seems to reach high values when the network is overloaded or traffic is unevenly distributed. The service delay of a channel is the sum of the waiting time-slots and the service time-slots encountered at the channel immediately following it.

3) A *traverse delay* ($d_{traverse}$) is an additional delay that represents the time spent by a flit in traversing the rest of the channels along the packet's path. The following assumptions are made: firstly, the length of the worm is longer than the diameter of the network, and secondly, there is no blocking at the last stage (destination). When the tail of the packet has left the injecting channel, the head of the message must have already arrived at the destination. In any case, the remaining packet will be received gradually in flits (one flit per time cycle). It will therefore take another $L-1$ time slots for the entire packet to be received at the destination, where $L$ depicts the *radix* of the MIN (the length of the path taken by the packet).

Thus, summarising all the above individual delays, the total delay of a packet injected at an $(i)$ arbitrary input can be written as:

$$d_{total,i} = d_{wait(i,1)} + d_{serv(i,j)} + L - 1 \qquad (7)$$

Consequently, the average overall packet delay over all processing nodes $\left(\overline{d_{total}}\right)$ is calculated by:

$$\overline{d_{total}} = \frac{\sum_i \overline{d_{total,i}}}{N} = \frac{1}{N} \cdot \sum_i (\overline{d_{wait(i,1)}} + \overline{d_{serv(i,j)}}) + \overline{L} - 1) \qquad (8)$$

where $N$ is the total number of inputs/outputs of the fabric, and $\overline{L}$ is the average length of the path distance (i.e. the total number of stages in the MIN).

In Equation (8), the factors $d_{wait(i,j)}$ and $d_{serv(i,j)}$ are particularly difficult to precisely identify using an analytical approach. This is because the service delay $d_{serv(i,j)}$ depends on the corresponding service times provided by the next channels. The calculation of the next service delay must be calculated in reverse order, that is, from the last stage to the first. Based on queueing theory, the most suitable model that represent the current MIN is the M/G/m model, which assumes independent arrivals from stage to stage. However, this assumption is not accurate in wormhole routing, because the flits' forwarding process between the inner stages is not an independent process. For these reasons, we reject the analytical approach in the current project and instead develop a simulation in order to estimate the performance metrics of the proposed system.

4. **Reliability of the proposed MIN**

In our reliability analysis of the current MIN, *p-t-p reliability* is selected as a reliability indicator. As discussed above, *p-t-p reliability* takes into consideration the probability of at least one fault-free path existing between an input-output pair. When a packet enters in a MIN, it receives the suitable 'routing address' (RA), according to the routing needs. The RA has as many bits as the number of stages. So, the path from input to output is given. There is no alternative way for a specific packet. An arbitrary fault-free path contains a total of $\log_2 N$ SEs' buffers in series. Therefore, the *total p-t-p reliability* $R_{MIN}$ of an arbitrary input-output pair can be expressed by Equation (9):

$$R_{MIN} = \prod_{i=1}^{\log_2 N} R_{(2X2)SE(k,i)} \qquad (9)$$

where $R_{(2X2)SE(k,i)}$ depicts the reliability of a buffer $k$ ($k = 0 \text{ or } 1$) belonging in SE of $i$ stage.

Let's suppose that each SE contains memories with $n_{lanes}$ parallel storage lanes and let's assume that the *reliability* of an arbitrary memory lane $(j)$ is $r_{j,flit}$, which represents the probability of a switch component being operational.

The *reliability* (i.e. the ability of a SE to overcome all unexpected circumstances) of a $k$ buffer (where $k$ is 0 or 1) and belonging in an ($i$) SE can be calculated as:

$$R_{(2x2)SE(k,i)} = \left(1 - \prod_{j=1}^{n_{lanes}}(1 - r_{j,flit})\right) \qquad (10)$$

The formulae (10), does not depict the overall reliability that a SE has, but presents the reliability that one of the buffers $k$ ($k = 0 \text{ or } 1$) of SE($i$) contributes in an input-to-output path.

On the other hand, when an arbitrary flit enters in a buffer, it has the ability to select one available lane. That makes constructions which are implemented with multiple parallel lanes have greater overall reliability compared to a corresponding system made by an exclusive one lane.

That results in the fabrics which are implemented with multiple parallel lanes to have greater overall reliability compared to a corresponding system made by an exclusive one lane. The buffer of parallel lanes seldom is occupied at all. During the operation when worms are starting gradually to be blocked, the parallel lines are progressively filled.

Combining Equations (9) and (10), the *overall reliability* of the MIN is given by:

$$R_{MIN} = \prod_{i=1}^{\log_2 N} \left(1 - \prod_{j=1}^{n_{lanes}}(1 - r_{j,flit})\right) \qquad (11)$$

If all the SEs in the MIN are identical, and if they have memories divided into the same number of segments, with the reliability of each lane equal to $r_{flit}$, then the *overall reliability* of the MIN is:

$$R_{MIN} = \left(1 - (1 - r_{flit})^{n_{lanes}}\right)^{\log_2 N} \quad (12)$$

The formulas (11) and (12) depicts the *overall input-output reliability of a MIN* in function to $r_{flit}$ and $n^{lanes}$, which are manufacturing features of a buffer. It is obvious the overall value depends on the geometric construction that exists each time and the quality of memory lanes.

Figure 5 shows the total input-output reliability versus the reliability ($r_{flit}$) of each storage lane of the SEs employed in the current novel fabric. Here, the memories of the SEs are considered to be organised with double storage lanes in each SE.

The graph shows a gradual increase in the *total p-t-p reliability* for MINs with radix 3, …,10. Networks have double-lane (DL) buffers. For low values of individual lane-storage reliability *r* (e.g. $r_{flit}$ =0.9) it can clearly be seen that the total p-t-p reliability of high-radix networks (e.g. radix 10 (N=1024)) is smaller than for a low-radix network (e.g. radix 8 (N=256)). When the factor $r_{flit}$ tends to one, all the fabrics, regardless of network radix, tend to have the maximum p-t-p reliability (a value of one).

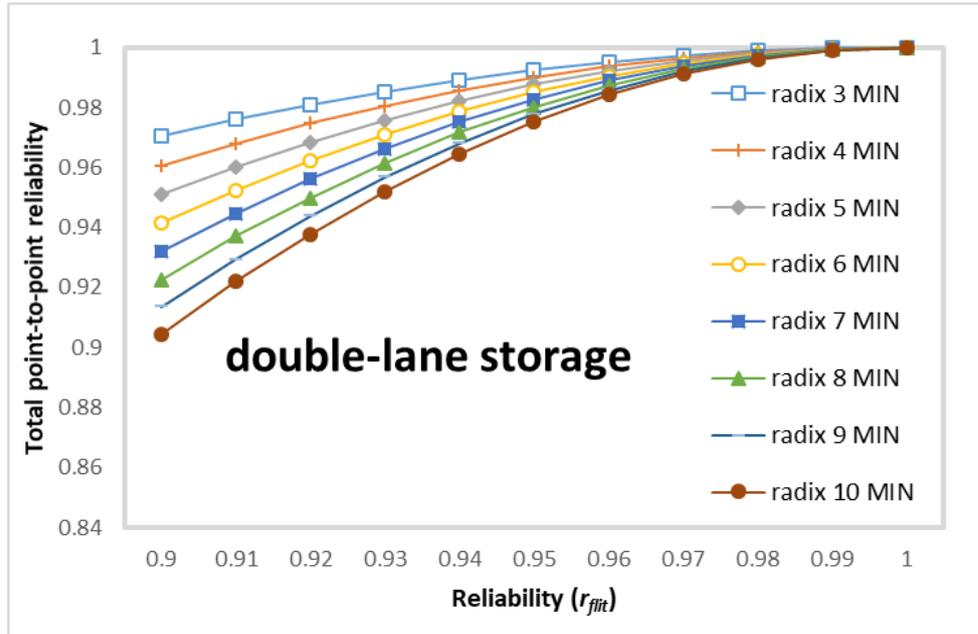

*Figure 5 Total path reliability versus lane-storage reliability ($r_{flit}$) for MINs with various values of radix and double-lane storage buffers ($n_{lanes} = 2$).*

Fig. 6 shows the total input-output reliability versus the reliability ($r_{flit}$) of each storage lane of SEs for MINs with various values of radix, for buffers with four-lane storage.

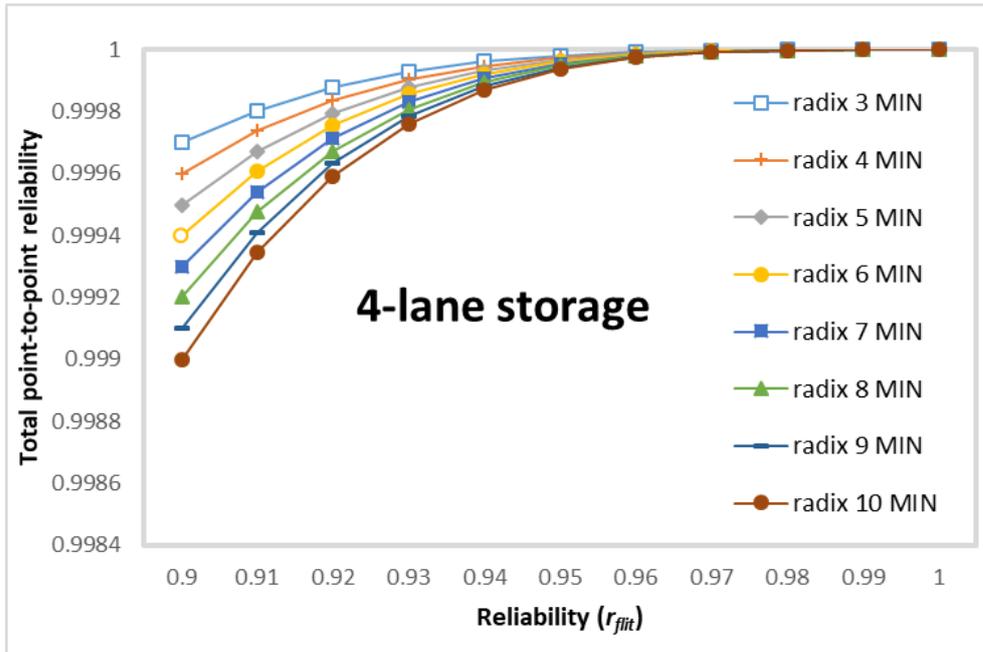

*Figure 6: Total path reliability of a MIN versus lane-storage reliability ( $r_{flit}$ ) for various network diameters composed of SEs with four-lane storage ($n_{lanes} = 4$) in each pair of inlets/outlets*

By comparing the corresponding values of the total path reliability (shown in Fig. 5 and 6), it becomes obvious that as the number of storage lanes in a MIN increases, the overall reliability improves rapidly, converging to the maximum possible value of one.

Fig. 7 depicts the total input-output reliability of MINs using single-lane (SL) and DL storage, respectively. The solid lines show the overall p-t-p reliability of MINs that use DL storage, while the dashed lines reveal the corresponding reliability of MINs that use SL storage (monolithic memory).

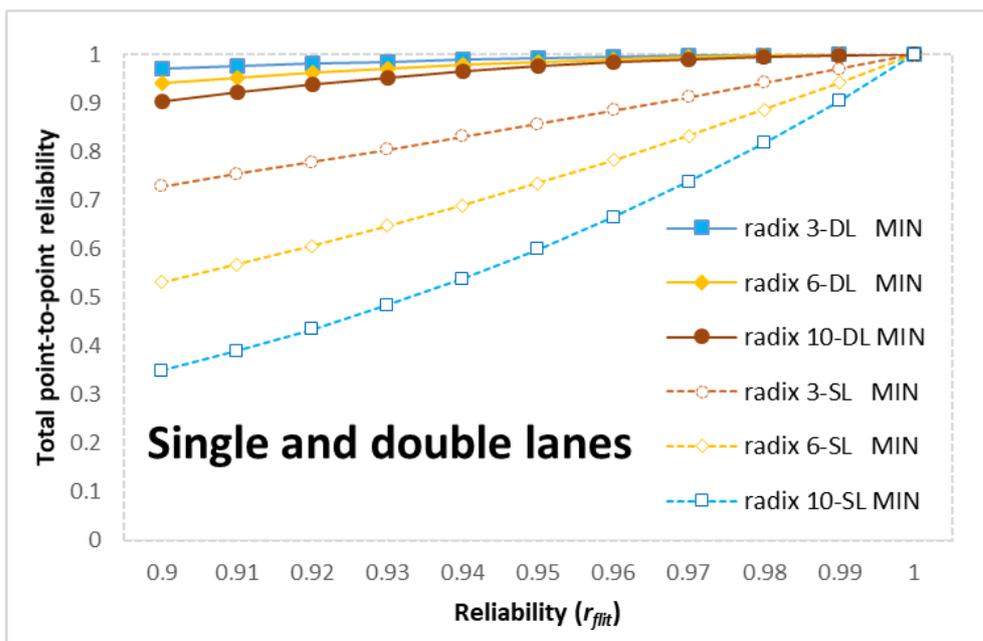

*Figure 7 Total path reliability of a MIN versus lane-storage reliability ( $r_{flit}$ ) for MINs with various values of radix, composed of SEs with single- and double-lane storages in each buffer*

The first observation is that MINs with multi-lane storage have values for overall reliability that are in the upper 10% of possible range, while MINs with SL storage have a very low level (as low as 0.3 in the case of an SL 10-radix MIN).

In brief, the fabric under study is very effective in terms of p-t-p reliability, and among other aspects this makes it a desirable approach.

Node: In figures 5, 6 and 7, the horizontal axis depicts the reliability of a typical lane which is a construction feature that is given by manufactures. The vertical axis depicts the overall point to point reliability which is an overall feature of the fabric. These are two different concepts.

## 5. Cooperative study of complexity and cost analysis

One of the chief issues in the field of MIN design is the cost, as an efficient system with high levels of cost offers little practical advantage. However, the cost of a MIN is in direct proportion to its *complexity*.

**Complexity of various types of MIN**

The complexity of the proposed fabric is in proportion to the number of SEs used in the integrated structure. The proposed $N \times N$ MIN consists of $k \times k$ SEs in $L = \log_k N$ parallel stages, and thus contains a total number of SEs of $\left(\frac{N}{k}\right) \times \log_k N$. The overall population of SEs that make up the construction is a direct indicator of the system's complexity, where a typical SE is considered to be the basic construction module. For this study, we select $k = 2$.

It is also evident that the number of flit lanes with which the MIN is equipped is an additional factor in the overall complexity of the multistage devices. We assume that the overall complexity is multiplied by the number of storage lanes in each SE. Thus, the overall complexity can be expressed by the following formula: $\left(\frac{N}{2}\right) \times \log_2 N \times n_{lanes}$ where $n_{lanes}$ depicts the number of storage lanes that are used in the memory of each SE.

In addition to the proposed MIN, other types of MIN have been selected for comparison in terms of their complexity/cost. The main reason for their selection were their relatively good performance and their high reliability of packet transport, as they are superior to classic MINs. The first column of Table I shows the types of MIN that are used for comparison. The characteristics of these MINs are relatively close to those of the proposed MIN. Some information is presented below for the selected MINs, from the existing literature.

- The **EGN** network was introduced by Wei and Lee [18]; this can maintain the permutation capability of a unique-path multistage network of size $N$ in the presence of any single fault.

- The **ASEN** network was studied thoroughly by Kumar et al. [19]. The main target of their research was to identify alternative paths when a switch failed to route at a particular stage.

- The **Pars MIN** architecture was introduced by Bistouni et al. [20]. The complexity function of the Pars MIN is given in their work.

The complexity functions of the above MINs are given in the second column of Table I below. All of the complexity formulae relate to fabrics with $2 \times 2$ SEs and $N$ inlets/outlets, with monolithic memories.

| Type of MIN | Complexity | Similarities | Differences |
|---|---|---|---|
| EGN | $6 \cdot N + 3 \cdot N((\log_2 N) - 1)$ | - $N \times N$ inlets | - Internal structure |
| ASEN | $6 \cdot N + \frac{9}{2} \cdot N \cdot ((\log_2 N) - 2)$ | - Store-and-forward routing | - Reliability |
| Pars | $6 \cdot N + 3 \cdot N \cdot ((\log_2 N) - 1)$ | | - Complexity |
| Two-layered MIN | $\frac{N}{2} \cdot ((\log_2 N) - 1) + 2 \cdot N$ | - Backpressure mechanism | - Cost |
| Three-layered MIN | $\frac{N}{2} \cdot ((\log_2 N) - 1) + 3 \cdot N$ | - Uniform traffic on inputs | |

*Table I Complexity of various types of MIN consisting of 2x2 SEs*

The **two-layered** and **three-layered MINs** in Table I refer to MINs with two and three layers (fan-out) at the last stage, respectively, which have been studied thoroughly by Garofalakis and Stergiou [21-22]. Their complexity functions are also given in Table I, accompanied by their similarities and differences.

**Costs of various types of MINs**

In the literature, several cost-effective approaches to wormhole routing have been presented for network schemas. Here, the approach of Bansal et al. [23, 24] is used to carry out our cost analysis. In this method, any $m \times n$ SE with SL storage in each inlet/outlet is considered to be an object with $m \times n$ cost units (e.g. a $2x2$ SE is considered as an object with four cost units). Now, assuming an apparatus is composed of SEs with DL storage per inlet/outlet, it therefore has cost units with double the value. This is completely reasonable because the construction of DL storage requires double the number of connections so that the final storage module can be efficiently controlled by the MIN's central control unit. Similarly, the cost scale is changed proportionally when the number of storage lanes in each memory module is increased.

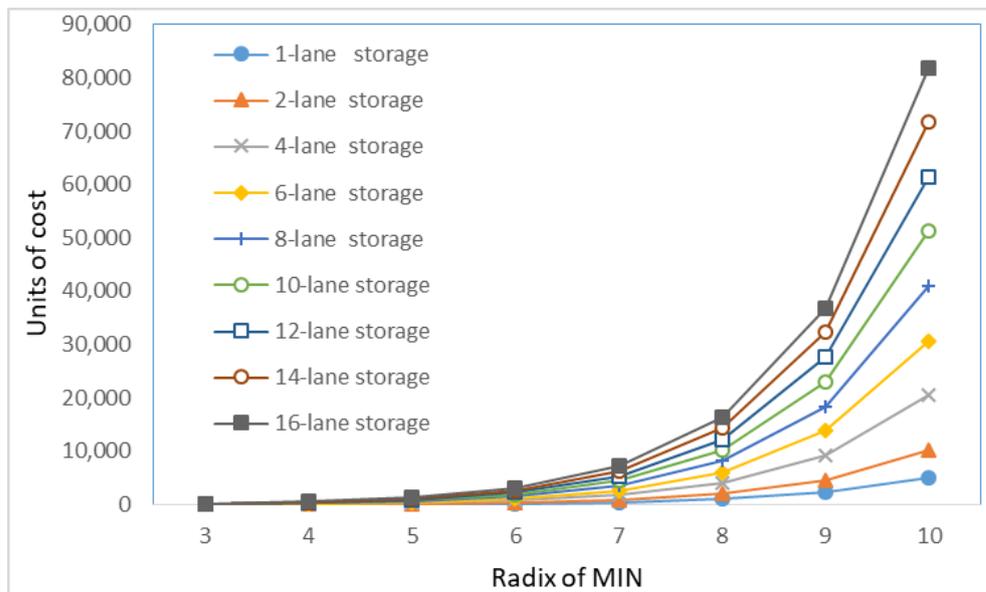

- *Figure 8 Units of cost versus radix of MINs with various numbers of storage lanes per buffer*

Using the above cost approach, cost calculations for various sizes of networks were performed. Fig. 8 displays the *units of cost* versus the radix of the MIN, for various numbers of storage lanes. For MINs with a radix less than six (64x64 inlets/outlets), the cost does not differ significantly and does not play such an important role. Above this value (a radix equal to or greater than six) the cost begins to differentiate and increases in proportion to the increase in the number of lanes.

This means that memory fragmentation into more than eight lanes ($4 \times 10^4$ units of cost in a 10-radix MIN) leads to expensive systems which are usually are ignored by the market, although technically they may operate effectively.

**Comparative plot of cost**

The overall values of unit of cost ($UoC$) are depicted in Fig. 9 for various types of MINs and the proposed MIN equipped with different numbers of storage lanes in the SEs.

The dashed lines represent the cost of the proposed MIN for different numbers of storage lanes in the memory (MIN-1L-B, MIN-2L-B, MIN-4L-B and MIN-10L-B). The solid lines depict other types of MIN, equipped with conventional SL memories but capable of providing excellent performance results. The lines for 2ML-1L-B and 3ML-1L-B represent the cost for a multi-layer MIN with two and three layers in the last stage, respectively [21-22].

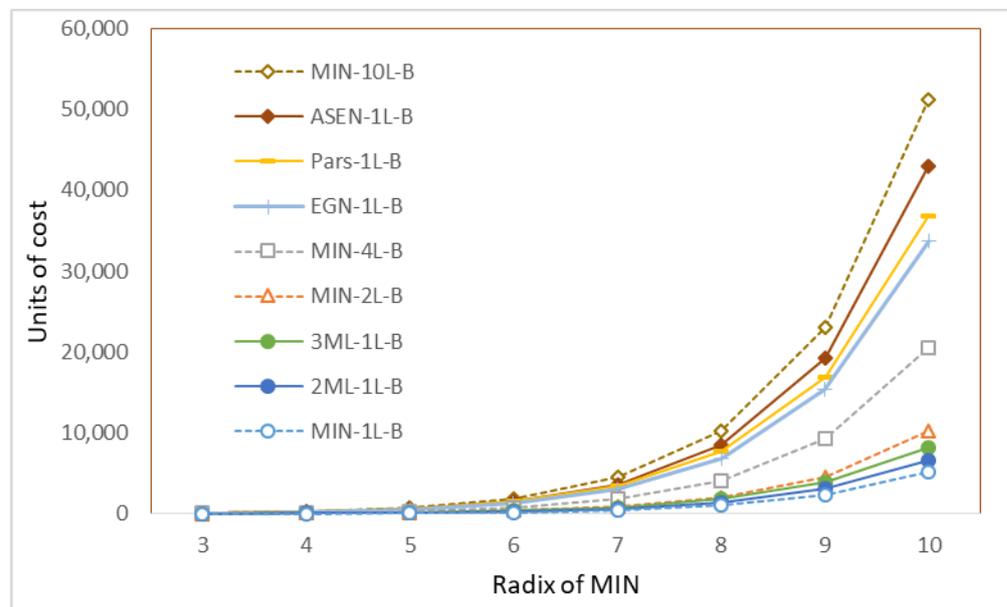

*Figure 9 Units of cost versus radix of MIN for various types of MIN equipped with single- or multi-lane memory*

We observe that the costs of the 2ML-1L-B and 3ML-1L-B MINs (solid lines) are lower than the proposed MIN with two-lane storage. On the other hand, the solid lines representing the Pars-1L-B, EGN-1L-B and ASEN-1L-B networks highlight the costs of Pars networks, which use redundant internal paths, and the EGN and ASEN types of multistage networks. These fabrics incur higher costs than the proposed MIN with two-lane storage, but provide satisfactory performance characteristics. However, the 10-lane MIN (the upper dashed lines) has a higher cost than the Pars, EGN and ASEN MINs.

The conclusion that emerges from this analysis is that the continuous fragmentation of memory into more segments causes corresponding increases in cost to high levels that are prohibited by the market. Thus, the designers of MINs can use the current cost-based approach to select the optimal number of memory lanes that should be implemented, not only for efficient overall MIN performance but also for cost-effectiveness.

## 6. The simulator

Here, we estimate the performance of the proposed MIN using simulations. We investigate $N \times N$ networks (for values of $N$ equal to 16, 64, 256 and 1024) consisting of $2x2$ SEs with internal queues that are organised into separate segments (lanes) and that use the wormhole routing technique. We developed a general simulator for these fabrics that was capable of handling several load conditions, operating at the flit level environment. The simulator was programmed in C and was capable of running various configuration schemas. When building the simulator, each $2x2$ SE was modelled by buffered queues matching the number of memory divisions (lanes). Each of these buffer's lane operates according to FCFS principles. All the flits are moved in a snaked manner, according to the routing tag of the header flit. Any cases of flit contention are solved randomly, with equal probability. As input parameters, we use the *probability of packet arrival*, the number of flit packets, the buffer length, the number of input/output ports and the number of stages.

Appendix A gives the main routine of the simulation, and shows the succession of steps involved in forwarding data flits. There are three main procedures: the *process of forwarding the header flit into the fabric*, the *process of forwarding the rest of the flits into the fabric* and the *process of forwarding flits out of the fabric*, which integrates the worm movement (the compact chain of flits) into the MIN.

In our simulation, in order to avoid unstable network operation, which arises at the beginning of operation of the network, the first 1,000-time cycles were ignored. In the steady state, we have a constant injection rate, which is equivalent to a constant number of packets (or equivalent flits).

Each packet is generated in the source queue of a corresponding input node, and is immediately inserted into the network when the selected virtual channel has enough space. The steady state situation is determined using the following approach: statistics are gathered for several mutually exclusive time periods that are relatively large. When the standard deviation of these statistics is within a small percentage (e.g. 4%) of the mean value, we consider that we have reached a steady state. Our termination criterion can be expressed as: minimum (steady state cycle, max cycle limit).

To improve the final accuracy of the simulation, several independent experiments are executed, and the average of these values is considered to be the final result. In cases where the network tends to saturation, all the statistics begin to oscillate, and longer simulation times are then required. When simulation results for different operating conditions are obtained, the data are normalised in order to enable a valid comparison. The next section presents the performance results from the simulator for the proposed fabric.

## 7. Performance results

**Performance metrics**

*Normalised throughput*

Figure 10 shows the *normalised throughput* versus the number of storage lanes implemented in a MIN operating with wormhole routing. The results are presented for network sizes with radix $x$ equal to 4, 5, 7, 9 and 10. The number of storage lanes in this study varied from one to 12, in steps of one. A conventional network is generated when the number of storage lanes is equal to one. The offered load is 80%, which is distributed uniformly, and the intermediate links are allocated randomly. The simulations are run with packets with at most 12 flits stored per channel.

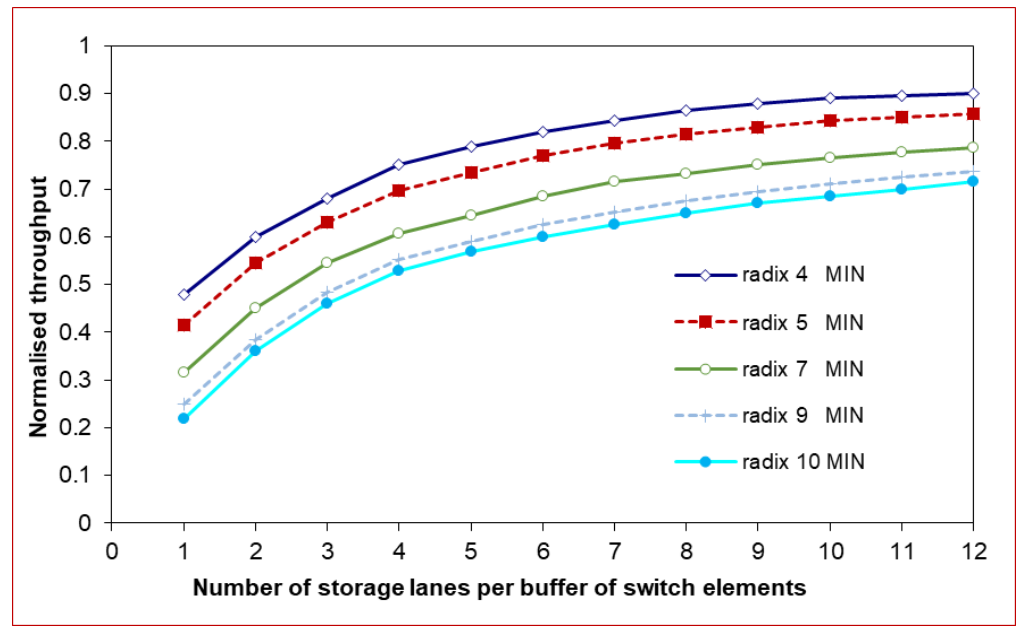

*Figure 10 Normalised throughput vs number of storage lanes per buffer for each switch element for MINs with various sizes of radix*

Fig. 10 shows that as the number of storage lanes increases, the corresponding throughput also increases. However, above approximately nine lanes, the value of throughput gradually reaches saturation. The plot reveals that the smaller the radix of the MIN, the better the throughput metric. Here, the best scenario is a network of radix four. This is due to the fact that large-scale networks give rise to a much larger number of conflicts.

The MIN of radix 10 (1024 inputs) working with DL memory in each SE achieves a normalised throughput of approximately 30%, a gain of about 14% compared with a corresponding fabric using a conventional (monolithic) type of memory. The same MIN using 12-flit storage per channel also has a normalised throughput metric of 71.2%, a gain of 49.2% compared with the corresponding conventional (one-lane) MIN.

*Packet delay*

The simulation results for packet delay are generated for scenarios using MINs that differ in terms of the network size and of the flit-storage lanes number, all of which use the wormhole routing technique. Fig. 11 shows the packet delay versus normalised offered load rate for MINs of radix eight, for memory organised into one, two, four, six, eight and 10 lanes.

The plot shows that for an offered load rate of less than 20%, the memory organisation has no effect. Above this value, the organisation of memory affects the packet delay. The plots clearly show that the use of conventional memory in a fabric means that the packet delay rises rapidly immediately after the offered load exceeds 20%. This abrupt transition in the delay values is shown in the figure by the left-hand dashed line. When lanes are added to the buffer, the plot reveals that the sharp increase in the packet delay shifts to increasingly high values of offered load rates, making the gradient of the corresponding curves more and more shallow.

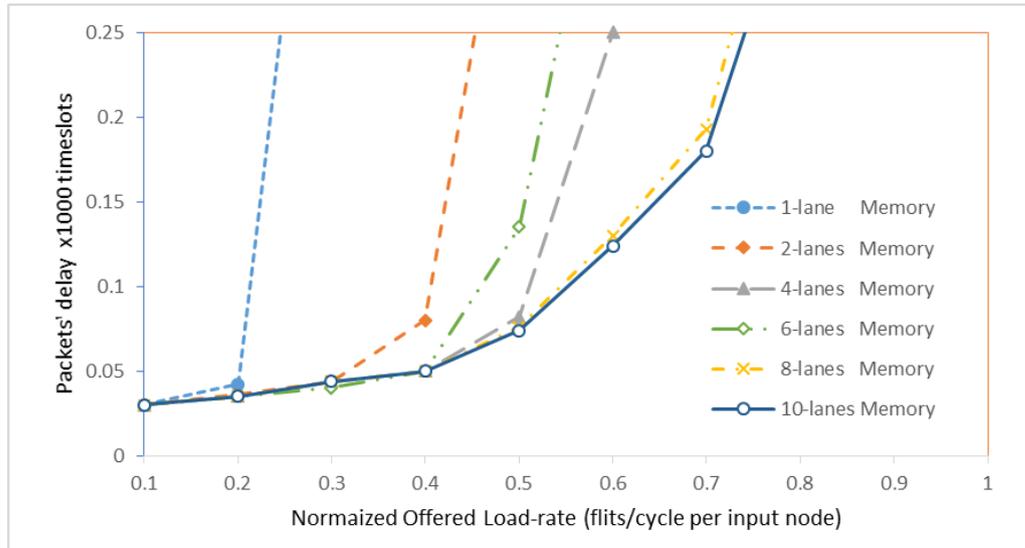

*Figure 11 Packet delay in timeslots versus various load rates for a MIN working with different memory lanes*

For an offered load of approximately 70%, a MIN with 10-lane memory shows a packet delay of 180 timeslots, while the delay in the corresponding MIN with conventional memory (one lane) reached infinity (practically in deadlock) at a much lower load. The experiments show that a delay of above 100–120 time cycles must be avoided if we have to handle voice traffic or video combined with voice data. This requirement is achieved if we use a MIN with 2–10 lanes and load it with less than 70% traffic, always using the wormhole routing method.

Due to the above presented behavior of multistage fabrics and especially the usage of wormhole mechanism has been included in many patents [25]. Moreover, the implementation of such systems in parallel processing technology provide more efficient systems and in general the method of wormhole routing is widely used in modern Massively Parallel Processing (MPP) systems. Wormhole routing technique requires fewer buffers and makes message latency insensitive to the distance that message traverses [26].

*Recent invention related to the performance of MINs*

The current detailed study of MINs concerns conventional electronic-type memories that usually work at the nanosecond scale. However, a form of memory recently invented by Tsakyridis et al. [27] has increased the data access speed to a speed of 10 Gbps. If this new approach is applied to MIN technology, this will result in the common interconnection apparatus moving into a new era in terms of data transfer rates. The usage of these new technologies has been already tested in the laboratory, giving excellent results. This latest invention allows electronic switching technology to operate beyond the nanosecond scale, at ultra-fast data rates.

## 8. Conclusion

The throughput of an interconnection network operating via wormhole routing can be increased by dividing the storage associated with each element into numerous memory lanes. Several small memory divisions can be associated with a physical channel, and this structure can be used instead of a single deep queue. The organisation of memory into lanes decouples buffer resources from the corresponding transmission resources. This technique allows packets to pass rapidly, better exploiting the capability of the network. Simulation studies show that for a fixed amount of buffer storage per link, memory segmentation increases the throughput many times, approaching the aptitude of the network. MIN designers and operators can use the results presented in this paper to select an optimal number of parallel queues, based on the QoS they want to offer to packets of different applications and the overall MIN performance they want to achieve.

Future work will focus on investigating other load configurations, including hot-spot, multi-priority and burst-type loads applied to the present MIN architecture with multi-lane storage. Moreover, by changing the buffers using new, more sophisticated and robust approaches but at the same time keeping the operating logic (e.g. using the packets and flits as basic data chunks, and using wormhole routing and memory organisation), the performance and reliability metrics are likely to be rapidly increased, and a new performance framework can be achieved. A new ultra-fast type of memory reported in [25] is a prime example. We therefore need to study multistage networks again in relation to new memory trends and from new viewpoints.

## Appendix A.  Algorithm

Main ()
{
Set *network size N*, *probability of packet arrivals (p)*,
 *number of flits per packet* ($n_{flits}$), *number of lanes per buffer* ($n_{lanes}$)

Initialise all stages
*Loop* = 1
**WHILE** *Loop* ≤ *N* **DO**
    *Process* Generate packets with a given probability
      of arrival for every inlet
    *Process* Split packets into flits
    /* inject flits into fabric inputs */
    *Process* Inject flits into first stage of MIN
 /*Forward worms */
 **FOR** *Stage* = 1 **TO** ($Log_2 N - 1$) **DO**
    **FOR** {*each buffer* of SE} **DO**
      **FOR** {each *lane storage*, ($n_{lanes}$)} **DO**
        **IF** (flit exists and flit is *header flit*, find next space) **{**
          *Process* Forward header flit into fabric
            **UNTIL** (the next space is eliminated)
          **WHILE** *number of flits* ≤ number of tail flits
            *Process* Forward the rest of flits into fabric
          **END WHILE** {tail flit, ($n_{flits}$)}
        **}** /*end IF *header flit**/
      **END FOR** {*lane storage*, ($n_{lanes}$)}
    **END** {*each buffer* of SE}
 **END FOR** {($Log_2 N - 1$) Stages}
 /* Last stage, export worms */
**FOR**   *Stage* = $Log_2 N$ **DO**
    **FOR** {*each buffer* of SE} **DO**
      **FOR** {*each lane storage*, ($n_{lanes}$)} **DO**
        **IF** (flit exists) **{**
          *Process* Forward flits out of fabric
            **UNTIL** (*number of flits* is tail flit)
          Collect data
        **}**
      **END FOR** {*lane storage*, ($n_{lanes}$)}
    **END** {*each buffer* of SE}
**END FOR** { *Stage* = $Log_2 N$ }
   *Loop* = *Loop* + 1
**ENDWHILE** {loop}
Gather and print the results
}

----- ------